\documentstyle[prd,aps,floats,epsfig]{revtex}

\begin{document}
\draft         
\title{Top Quark Pair Production at $e^+e^-$ Colliders
in the Topcolor-assisted Technicolor Model
\thanks{Supported by National Natural Science Foundation of China under
Grant No.19775012, and Natural Science Foundation of Henan Educational
Committee under No.1999140013.}}
\author{HUANG Jinshu$^1$, XIONG Zhaohua$^2$, and LU Gongru$^3$}
\address{$^1$ Department of Physics, Henan Nanyang Normal College, Nanyang
473061, China}
\address{ $^2$ Institute of Theoretical Physics, Academia Sinica, Beijing
100080, China}
\address{$^3$ Department of Physics, Henan Normal University, Xinxiang
453002, China}
\date{\today}
\maketitle
 
\maketitle

\begin{abstract}
In the framework of topcolor-assisted technicolor model we calculate the
contributions from the pseudo Goldstone bosons and new gauge bosons to
$e^+e^- \rightarrow t\bar{t}$. We find that, for reasonable ranges of the
parameters, the pseudo Goldstone bosons afford dominate contribution, the
correction arising from new gauge bosons is negligibly small, the maximum 
of the relative corrections is $-10\%$ with the center-of-mass energy 
$\sqrt{s}=500$ GeV; whereas in case of $\sqrt{s}=1500$ GeV, the relative
corrections could be up to  $16\%$. Thus large new physics might be observable at
the experiments of next-generation linear colliders.
\end{abstract}

\pacs{PACS: 12.60.Nz, 13.40.-f, 14.65.Ha}

{\bf Key words:} pseudo Goldstone boson, gauge boson, technicolor, top quark.

\section{Introduction}
It is often argued that the standard model (SM) should be augmented
by new physics at higher energy scales because of some unanswered 
fundamental questions. The recently reported 2.6 standard deviation of the 
muon anomalous magnetic moment over its SM prediction \cite{E821} may serve 
as the first evidence of existence of new physics at a scale not far above 
the weak scale. There are numerous speculations on the possible forms of new
physics, among which supersymmetry and technicolor are the two typical
different frameworks.

Technicolor---a strong interaction of fermions and gauge bosons at
the scale $\Lambda_{\rm TC} \sim 1\ {\rm TeV}$--- is a scenario for the
dynamical breakdown of electroweak symmetry to electromagnetism
\cite{Weinberg79}. Based on the similar phenomenon of chiral symmetry
breakdown in QCD, technicolor is explicitly defined and completely
natural. To account for the mass of quarks, leptons, and Goldstone
``technipions'' in such a scheme, technicolor(TC), ordinary color, and flavor
symmetry are embedded in a large gauge group, called extended technicolor
(ETC) \cite{Dimopoulos79}. Because of the conflict between constraints
on  flavor-changing neutral  currents (FCNC) and the magnitude of
ETC-generated quark, lepton and technipion masses, classical technicolor
was superseded by a ``walking'' technicolor and ``multiscale technicolor''
\cite{Holdom81,Lane89}. The incapability of explain the top quark's
large mass without afoul of either cherished  notions of naturalness
or experiments from the $\rho$ parameter and the ${\em Z}\rightarrow
b\bar{b}$ decay rate  by ETC \cite{Chivukula92} led to the
topcolor-assisted technicolor \cite{Miransky89} and  the technicolor 
with scalars model \cite{Simmons89,Xiong01}.

Up to now, top quark is the heaviest particle discovered in experiments.
Its mass, $m_t=175$ GeV \cite{Abe95}, is of the order of the electroweak 
spontaneous breaking (EWSB) scale
which means the top quark couples rather strongly to the EWSB
sector. So effects from new physics would be more apparent in processes
with the top quark than with any other light quarks. On the experimental side,
it is possible to separately measure various production and decay form
factors of the top quark at the level of a few percent \cite{Peskin92}.
Therefore, theoretical calculations of various corrections to the production
and decay of the top quark are of much interest. 

Top quark pair can be produced at various high energy colliders. 
This paper is devoted   to examine the ability of the suggested future TeV 
energy $e^+e^-$ colliders in testing TC effects via $t\bar{t}$ production. 
It is organized as follows. In Sec.~\ref{sec:TC}, we present a brief
review of the original topcolor-assisted technicolor (TOPCTC) model by
C. T. Hill \cite{tctwohill}. In Sec.~\ref{sec:eett}, we give out the 
corrections 
to the $e^+e^- \rightarrow t\bar{t}$ cross section at the center-of-mass
(c.m.) energy $\sqrt{s}=0.5$, $1.0$ and $1.5$ TeV colliders in the TOPCTC 
model. Both the effects of pseudo Goldstone bosons (PGBs) and new gauge 
bosons are evaluated. Conclusions are 
included in Sec.~\ref{sec:summary}. The analytic formulas for the form 
factors in the production amplitudes are presented in the Appendix.

\section{Topcolor-assisted technicolor model}
\label{sec:TC}
The model assumes \cite{topcref,tctwohill,tctwoklee}: (i) electroweak
interactions are broken by technicolor; (ii) the top quark mass is large
because it is the combination of a dynamical condensate component
$(1-\varepsilon)m_t$, generated by a new strong dynamics, together with a
small fundamental component $\varepsilon m_t (\varepsilon \ll 1)$, generated
by ETC; (iii) the new strong dynamics is assumed to be chiral critically
strong but spontaneously broken by technicolor at the scale $\sim 1~{\rm
TeV}$, and it generally couples preferentially to the third  generation.
This needs a new class of technicolor models incorporating ``top-color''
(TOPC). The dynamics at $\sim 1~{\rm TeV}$ scale involves the gauge structure:
\begin{equation}
 SU(3)_1\times SU(3)_2\times U(1)_{Y_1}\times U(1)_{Y_2}
\rightarrow SU(3)_{\rm QCD}\times U(1)_{EM} \nonumber
\end{equation}
where $SU(3)_1\times U(1)_{Y_1}~[SU(3)_2\times U(1)_{Y_2}]$ generally
couples preferentially to the third (first and second) generation, and is
assumed to be strong enough to form chiral $<\bar{t}t>$ but not $<\bar{b}b>$
condensation by the $U(1)_{Y_1}$ coupling.  A residual global symmetry
$SU(3)' \times U(1)'$ implies the existence of a massive color-singlet
heavy $Z'$ and an octet $B_\mu^A$. A symmetry-breaking pattern onlined
above will generically give rise to three top-pions, $\pi_{t}$, near the
top mass scale.

The new interactions by the topcolor for the process
$e^+e^- \rightarrow t\bar{t}$ give 
\begin{eqnarray}
Z'_\mu e^+e^-:&& g_1\tan\theta'\gamma_{\mu}\left[\frac{1}{2}L+R\right],
\nonumber\\
Z'_\mu t\bar{t}:&&g_1\cot\theta'\gamma^{\mu}\left[\frac{1}{6}L+
\frac{2}{3}R\right],
\end{eqnarray}
where $R(L)=(1\pm\gamma_5)/2$, $g_1=\alpha_{\rm EM}/\cos \theta_{W}$ is
the $U(1)_Y$ coupling constant at the scale $\sim 1~{\rm TeV}$
with $\theta_W$ being the Weinberg angle. The SM
$U(1)_Y$ and the $U(1)'$ field $Z_\alpha'$ are then defined by
orthogonal rotation with mixing angle $\theta'$. 

There exists the ETC gauge bosons including the sideways and diagonal gauge
bosons $Z^*$ in this model. The coupling of $Z^*$ to the fermions and
technifermions can be found in Ref. \cite{Wu95}. For the sake of simplicity,
we assume that the mass of the sideways gauge boson is equal to the mass of
the diagonal gauge boson, namely $m_{Z^*}$, so the $Z^*_\mu e^+e^-$ and 
$Z^*_\mu t\bar{t}$ vertices by the ETC dynamics can be written as
\begin{eqnarray}
Z^*_\mu e^+e^-:&&
-\frac{\varepsilon m_t}{16\pi f_{\pi}} \frac{e}{s_Wc_W}\gamma_{\mu}
\left\{\left[\frac{2N_C}{N_{TC}+1}\xi_t^{-1}(1-\xi_e \xi_{\nu})\xi_e
-(10\varepsilon)^{-2/3}\xi_e^2\right]L-
\left[\frac{2N_C}{N_{TC}+1}(\xi_e\xi_{\nu})\xi_t^{-1}
(1-\xi_e \xi_{\nu})\xi_e^{-1}\right]R\right\},\nonumber \\
Z^*_\mu t\bar{t}:&& -\frac{\varepsilon m_t}{16\pi f_{\pi}} \frac{e}{s_Wc_W}
\sqrt{\frac{N_{TC}}{N_C}}\gamma_{\mu}
\left[\left(\frac{2N_C}{N_{TC}+1}+\xi_t^2\right)L-
\left(\frac{2N_C}{N_{TC}+1}-1\right)\xi_t^{-2}R\right]
\end{eqnarray}
where $N_{TC}$ and $N_C$ are the numbers of technicolors and ordinary colors,
respectively; $s_W = \sin \theta_W$, $c_W = \cos \theta_W$; 
$\xi_t$, $\xi_e$ and $\xi_{\nu}$ are coupling
coefficients and are ETC gauge-group dependent. Following Ref. \cite{Wu95},
we take $\xi_t=\xi_e=1/\sqrt{2}$, $\xi_{\nu}=0.1 \xi_e^{-1}$.

In this TOPCTC model, there are 60 technipions in the ETC sector with
decay constant $f_{\pi}=123$ GeV and three top pions $\pi_t^0, \pi_t^{\pm}$
in the TOPC sector with decay constant $f_{\pi_t}=50$ GeV. The ETC
sector is one generation technicolor model. The relevant technipions in 
this study are only the color-singlet $\pi$ and color-octet $\pi_8$.
 The color-singlets $\pi$ include the isosinglet scalar $\pi^0$, the
isotriplet scalar ($p^{\pm}$, $p^{3}$), while the color-octets $\pi_8$
involve the isosinglet $p_8^0$ and  the iso-octet scalar $p_8^\pm,~p_8^3$. 
The color-singlet(octet) technipion-top(bottom) interactions are given by
\begin{eqnarray}
&&\frac{c_t \varepsilon m_t}{\sqrt{2}f_{\pi}}\left[i\bar{t}\gamma_5t\pi^0
+i\bar{t}\gamma_5 t\pi^3+ \frac{1}{\sqrt{2}}\bar{t}(1-\gamma_5)b\pi^+
+\frac{1}{\sqrt{2}} \bar{b}(1+\gamma_5)t\pi^-\right],\nonumber \\
&&\frac{\sqrt{2} \varepsilon m_t}{f_{\pi}}\left[i\bar{t}\gamma_5
\frac{\lambda^a}{2}
t\pi_8^0 +i\bar{t}\gamma_5\frac{\lambda^a}{2}t\pi_8^3+ \frac{1}{\sqrt{2}}
\bar{t}(1-\gamma_5)\frac{\lambda^a}{2}b\pi_8^+ +\frac{1}{\sqrt{2}}\bar{b}
(1+\gamma_5)\frac{\lambda^a}{2}t\pi_8^-\right],
\end{eqnarray}
where the coefficient $c_t=1/\sqrt{6}$,  and $\lambda^a$ is a Gell-Mann
matrix acting on ordinary color indices. 

The interaction of the top pions with  the top (bottom) quark has form
\begin{equation}
\frac{(1-\varepsilon)m_t}{\sqrt{2}f_{\pi_t}}\left[i \bar{t}\gamma_5 t
\pi_t^0
 +\frac{1}{\sqrt{2}}\bar{t}(1-\gamma_5)b\pi_t^+
 +\frac{1}{\sqrt{2}}\bar{b}(1+\gamma_5)t\pi_t^-\right].
\end{equation}
More detail Feynman rules needed in the calculations can be found in
Refs. \cite{Miransky89,Ellis81}.

\section{The $e^+e^- \rightarrow t\bar{t}$ cross section}
\label{sec:eett}

\subsection{The PGBs contributions}

The relevant Feynman diagrams arising from the PGBs 
contributing to the $e^+e^- \rightarrow t\bar{t}$ production amplitudes
are shown in Fig.~1(a)-(g).
\begin{figure}[hb]
\begin{center}
\psfig{file=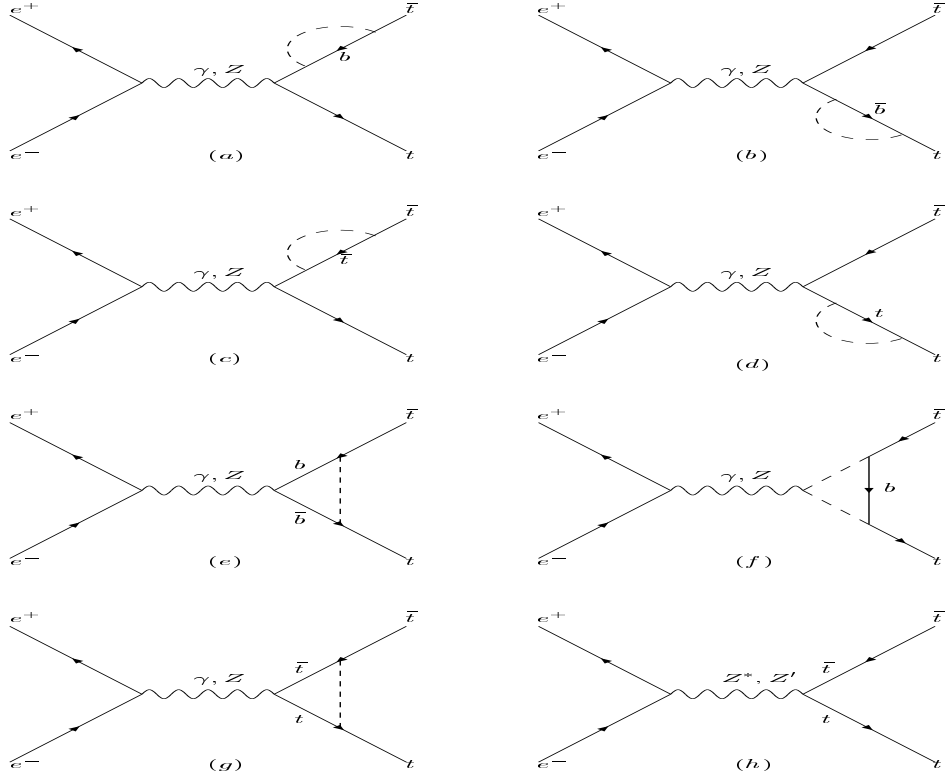,width=400pt,height=300pt}  
\caption{Feynman diagrams for PGB and new gauge boson contributions to the
$e^+e^- \rightarrow t\bar{t}$ process: (a-d) self-energy diagrams, 
(e-g) vertex diagrams, (h) the corrective diagram of
the gauge boson $Z^*$ and $Z'$. The dashed lines represent the charged
technipions $\pi^{\pm}$, $\pi_8^{\pm}$ and top pions $\pi_t^{\pm}$ in the
the diagrams (a,b,e,f), and the neutral technipions $\pi^0$, $\pi^3$,
$\pi_8^0$, $\pi_8^3$ and top pion $\pi_t^0$ in the diagrams (c,d,g).}
\end{center} 
\end{figure}              

In our calculation, we use the dimensional regularization 
to regulate all the ultraviolet divergences in the virtual loop corrections, 
and adopt the Feynman gauge and on-mass-shell renormalization scheme 
\cite{Bohm86}. The renormalized amplitude for $e^+e^- \rightarrow t\bar{t}$ 
contains
\begin{equation}
M=M^{\rm tree}+\delta M,
\end{equation}
where the tree-level amplitude
\begin{eqnarray}
M^{\rm tree}&=&\bar{u}(t)ie\gamma^{\mu}[v_t(1+\gamma_5)
+a_t(1-\gamma_5)]v(\bar{t})
\frac{-ig_{\mu \nu}}{(P_t+P_{\bar{t}})^2-m_Z^2}\bar{v}(e^+)ie\gamma^{\nu}
[v_e(1+\gamma_5)\nonumber\\ 
&&+a_e(1-\gamma_5)]u(e^-)+\bar{u}(t)i\frac{2}{3}e\gamma^{\mu}v(\bar{t})
\frac{-ig_{\mu\nu}}{(P_t+P_{\bar{t}})^2}\bar{v}(e^+)(-ie)\gamma^{\nu}u(e^-).
\end{eqnarray}
Here $P_{t,\bar{t}}$ denote the momentum of the outgoing top quark
pair. $\delta M$ represents the PGBs one-loop corrections which contains
\begin{eqnarray}
\delta M&=&\delta M_Z +\delta M_{\gamma}, \\
\delta M_Z&=&\bar{u}(t)\Gamma_Z v(\bar{t})\frac{-ig_{\mu \nu}}{(P_t
+P_{\bar{t}})^2-m_Z^2}\bar{v}(e^+)ie\gamma^{\nu}[v_e(1+\gamma_5)
+a_e(1-\gamma_5)]u(e^-), \\
\delta M_{\gamma}&=&\delta M_Z|_{Z \rightarrow \gamma, m_Z=0,
v_t=a_t=1/3,v_e=a_e=-1/2}
\end{eqnarray}
where $v_t=\frac{-(4/3)s_W^2}{4s_Wc_W}, \
a_t=\frac{1-(4/3)s_W^2}{4s_Wc_W}, 
v_e=\frac{2s_W^2}{4s_Wc_W}, \ 
a_e=\frac{1-2s_W^2}{4s_Wc_W}$. $\Gamma_Z, \Gamma_{\gamma}$ denote 
the effect vertex $Zt\bar{t}$ and $\gamma t\bar{t}$ arising from PGBs 
corrections, respectively, and  
\begin{eqnarray}
\Gamma_Z&=&ie[\gamma^{\mu}LF_{1Z} +\gamma^{\mu}RF_{2Z} +P_t^{\mu} L F_{3Z
}
+P_t^{\mu} R F_{4Z}+ P_{\bar{t}}^{\mu} L F_{5Z}
+P_{\bar{t}}^{\mu}R F_{6Z}], \\
\Gamma_{\gamma}&=&\Gamma_Z|_{F_{iz} \rightarrow F_{i\gamma}}.
\end{eqnarray}
The form factors $\Gamma_{iZ}$, $\Gamma_{i\gamma}$ expressed in terms
of two- and three-point scalar integrals
are presented in the Appendix. It is easy to find that all the ultraviolet
divergences cancel in the effective vertex.

Now we express the differential cross section of the process 
$e^+e^- \rightarrow t\bar{t}$ as 
\begin{equation}
{\rm d}\sigma= \frac{(2\pi)^4\delta^4(P_{e^+}+P_{e^-}-P_t-P_{\bar{t}})}
{4\sqrt{(P_{e^+}P_{e^-})^2-(m_{e^+}m_{e^-})^2}}\bar{\sum}|M|^2
\frac{{\rm d}^3\vec{P}_{t}}{(2\pi)^32E_{t}} \frac{{\rm d}^3\vec{P}_{\bar{t}}}
{(2\pi)^32E_{\bar{t}}}.
\label{dsigma}
\end{equation}
Integrating out phase space, we obtain the total cross section 
\begin{equation}
\sigma=\frac{\sqrt{s(s-4m_t^2)}}{32\pi s^2}\int_{-1}^{1}\bar{\sum}|M|^2
{\rm d\ cos}\theta=\sigma_0 + \delta \sigma,
\end{equation}
where $\sigma_0$ is the cross section at the tree-level, and $\delta \sigma$ 
denotes the Yukawa corrections arising from PGBs at the one-loop level.

Through the above equations, we calculate the relative corrections to the
production rate $\sigma$ arising from PGBs. Since the ETC sector of this
model is one generation technicolor model, the masses of the PGBs are model
dependent. In Ref. \cite{Ellis81}, the masses for  $\pi$ and 
$\pi_8$ are taken to be in the range $60\ GeV< m_{\pi}< 200\ GeV$, 
$200\ GeV< m_{\pi_8}< 500\ GeV$. In the TOPC sector, the mass of the
top-pion, $m_{\pi_t}$, a reasonable value of the parameter is around 200
GeV. In the following calculation,  we would rather take a slightly larger
range, $150\ GeV< m_{\pi_t}< 450\ GeV$, to see its effect, and
shall take the masses of $m_{\pi}$, 150 GeV, and $m_{\pi_8}$, 246 GeV.
Furthermore, the common mass $m_\pi$ is assumed for
color-singlets and $m_{\pi_8}$ for colore-octet scalars. 
The final numerical results are plotted in Figs.2 and 3.

\begin{figure}[htb]
\begin{center}
\epsfig{file=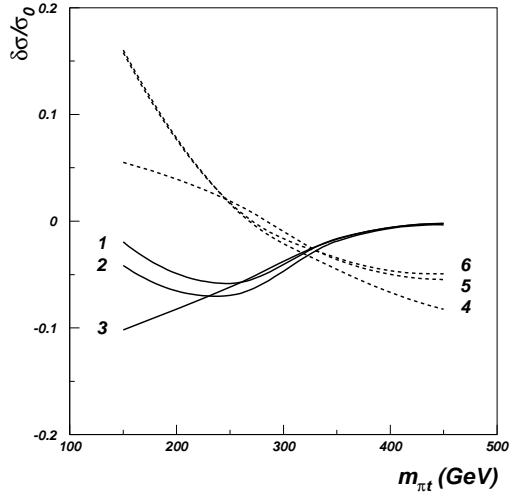,width=7cm}
\caption{The relative correction $\delta \sigma/\sigma_0$ curves as a
function of $m_{\pi_t}$. The solid lines 1, 2 and 3 correspond to 
$\varepsilon= 0.03, 0.07, 0.1$ with $\sqrt{s}=500$ GeV, respectively, 
while the dashed ones 4,5 and 6,  with $\sqrt{s}=1500$ GeV.}
\end{center}    
\end{figure}              

Figure 2 shows $\delta \sigma/ \sigma_0$ versus
$m_{\pi_t}$.  When $\sqrt{s}=500$ GeV, one can see that (i) in case of 
$m_{\pi_t}>250$ GeV, the size of the relative corrections $\delta\sigma/
\sigma_0$ decrease near to zero with $m_{\pi_t}$ sensitively, 
(ii) the maximum of the relative corrections can reach $-10\%$ 
for $\varepsilon=0.07$, $m_{\pi_t}=150$ GeV. However, 
when $\sqrt{s}=1500$ GeV,  $\delta \sigma/ \sigma_0$ changes from positive 
to negative, and negative correction becomes larger when the top-pion is 
heavier. In this case,  the relative corrections can be up to  $16\%$ 
when $m_{\pi_t}=150$ GeV, $\varepsilon=0.03$ or $0.1$. 
\begin{figure}[htb]
\begin{center}
\epsfig{file=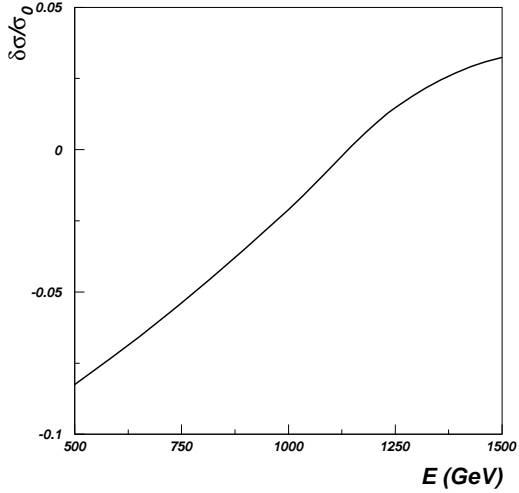,width=7cm}
\end{center}             
\caption{$\delta \sigma/\sigma_0$ as a function of  $E=\sqrt{s}$
with $m_{\pi_t}=220$ GeV, $\varepsilon=0.07$.}
\end{figure} 

The dependence of the relative correction on $\sqrt{s}$  is shown in Fig. 3.
The negative contributions from PGBs to
$e^+e^- \rightarrow t\bar{t}$ are found at the outset stage, 
but at $\sqrt{s}=1150$ GeV, the effects become positive.

\subsection{The gauge bosons corrections}

Now let's consider the contributions from the new gauge boson to
the $e^+e^- \rightarrow t\bar{t}$ cross section.

In the TOPCTC theory, there are two kinds of new gauge bosons: the ETC
gauge bosons $Z^*$ including the sideways and diagonal gauge bosons, and 
the TOPC gauge bosons including the color-octet colorons $B_{\mu}$ and
color-singlet $Z'$. The relevant gauge bosons to the process $e^+e^-
\rightarrow t\bar{t}$ are only $Z^*$ in ETC sector and $Z'$ in TOPC sector,
the Feynman diagram is shown in Fig.~1(h).

Using formula (2,3) we can obtain the effects of  the gauge bosons. 
The contributions from ETC gauge boson to the $e^+e^- \rightarrow
t\bar{t}$ cross section are found to incease quickly with $\varepsilon$ and
$\sqrt{s}$, and decrease slowly with $m_{Z^*}$, ans 
the maximum of the relative
corrections $\delta \sigma_{Z^*}/\sigma_0$ is only the order of $10^{-12}$ 
whatever $\varepsilon$, $\sqrt{s}$ and $m_{Z^*}$ are taken in the above 
parameter ranges, therefore, can be neglected safely. 

\begin{figure}[hb]
\begin{center}
\epsfig{file=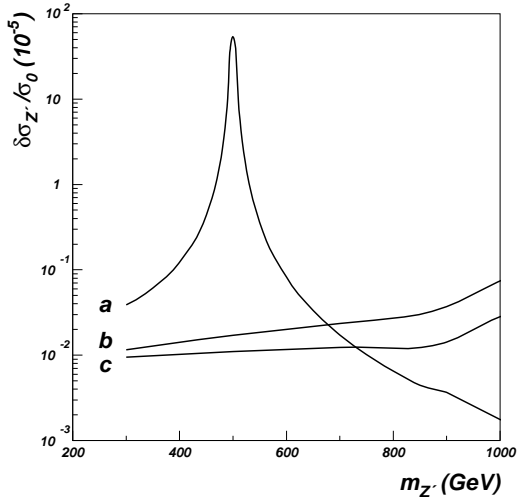,width=7cm}
\end{center}             
\caption{The relative
correction $\delta \sigma/\sigma_0$ of the gauge boson $Z'$
as a function $m_{Z'}$. The lines a, b and c correspond to 
$\sqrt{s}=500$, $1000$ and $1500$ GeV, respectively.}
\end{figure} 
 
As  for the contributions from $Z'$, it is obvious the contributions to the
$e^+e^- \rightarrow t\bar{t}$ cross section aren't related to $\theta '$
(see Eq. (2)). In our calculation, we assume
the mass of the gauge boson $Z'$ varying from 300\ Gev to 1000 GeV to 
study the effects of $Z'$.  The numerical results are plotted in Fig. 4. 
From this figure we find that
 except for the $Z^\prime$ resonance region, the relative correction 
$\delta \sigma_{Z'}/\sigma_0$ is also quite small, and  
 the corrections increase with $m_{Z'}$ 
at the outset stage and decrease at the back stage at
$\sqrt{s}=500$ and always increase in case of  $\sqrt{s}=1000, 1500$ GeV.

Now we would like extend this study to the top-color-assisted 
multiscale technicolor (TOPCMTC) model\cite{tctwoklee,Yue97}.
The main difference from the original TOPCTC model\cite{Miransky89} is the 
ETC sector.  Instead of the one generation technicolor model with 
$f_{\pi}=123$ GeV, $c_t=1/\sqrt{6}$ and $N_{TC}=4$ in Eq. (3-5) 
at the ETC sector in the original TOPCTC model, in TOPCMTC model, 
the ETC sector is the multiscale walking technicolor model with
$f_{\pi}=40$ GeV, $c_t=2/\sqrt{6}$ and $N_{TC}=6$,  
the corrections are similar to those of the original TOPCTC model.
The contribution from PGBs in the TOPCMTC model is slightly larger, 
the contribution from the gauge boson $Z\prime$ is the same
as that of the original TOPCTC model, and the correction from the ETC gauge
boson $Z^*$ is still negligibly small.

\section{The discussions and conclusions}
\label{sec:summary}

We have studied the contributions from the pseudo
Goldstone bosons and new gauge bosons in the original top-color-assisted
technicolor model to $t\bar{t}$ production at the $\sqrt{s}=0.5$, $1.0$,
and $1.5$ TeV $e^+e^-$ colliders. We found for reasonable ranges of the 
parameters, the pseudo Goldstone bosons afford
dominate contribution, the correction arising from new gauge bosons is
negligibly small, the maximum of the relative corrections is $-10\%$ with
the c.m. energy $\sqrt{s}=500$ GeV; whereas in case of $\sqrt{s}=1500$ 
GeV, the maximum of the relative corrections may reach $16\%$. 
We thus conclude that the $e^+e^- \rightarrow t\bar{t}$ experiments at the
future colliders are really interesting in testing the standard model and
searching for the signs of technicolor.

\appendix
\section*{Appendix}

The form factors $\Gamma_{iZ}$,$\Gamma_{i\gamma}$ of the effect vertex
$Zt\bar{t}$ and $\gamma t\bar{t}$ arising from PGBs corrections 
are given out
$$
F_{1Z}=\frac{1}{16\pi^2}\sum_{i=\pi, \pi_8, \pi_t}\lambda_i
\{ [-2a_t(B_1(-P_t,m_b,m_i)+ m_t^2B_1'(-P_t,m_b,m_i))-2kC_{24}^2
\ \hspace{1.2cm} \
$$
$$
+2v_b(-B_0(\sqrt{s},m_b,m_b)-m_i^2C_0^1+2C_{24}^1-m_t^2C_{11}^1)]
\ \hspace{3.6cm} \
$$
$$
+a_i[2a_tm_t^2(2C_0^3+C_{11}^3)+v_t(2C_{24}^3-B_0(-\sqrt{s},m_t,m_t)
-B_1(-P_t,m_t,m_i))
$$
$$
+v_tm_t^2(-2C_{11}^3+2C_{12}^3-C_0^3-2B_1'(-P_t,m_t,m_i)
-2B_0'(-P_t,m_t,m_i))] \} \ \ \
\eqno(a1)
$$
$$
F_{2Z}=\frac{1}{16\pi^2}\sum_{i=\pi, \pi_8, \pi_t}\lambda_i
\{ 2m_t^2[v_tB_1'(-P_t,m_b,m_i)+v_b(C_0^1+C_{11}^1)]
+ a_i [2v_tm_t^2(2C_0^3+C_{11}^3)
$$
$$
+a_t(2C_{24}^3-B_0(-\sqrt{s},m_t,m_t)
-B_1(-P_t,m_t,m_i))+a_t m_t^2(-2C_{11}^3+2C_{12}^3
$$
$$
-C_0^3-2B_1'(-P_t,m_t,m_i)-2B_0'(-P_t,m_t,m_i))]\} \ \hspace{4cm} \
\eqno(a2)
$$
$$
F_{3Z}=\frac{1}{16\pi^2}\sum_{i=\pi, \pi_8, \pi_t}\lambda_i \{
[4v_bm_t(C_{12}^1+C_{22}^1)+km_t(-C_{11}^2+C_{12}^2-2C_{21}^2-2C_{22}^2
+C_{23}^2)] \ \
$$
$$
+a_i[2v_tm_tC_{22}^3-2a_tm_t(C_{22}^3-C_{23}^3) ] \}
\ \hspace{6.1cm} \
\eqno(a3)
$$
$$
F_{4Z}=\frac{1}{16\pi^2}\sum_{i=\pi, \pi_8, \pi_t}\lambda_i \{
[4v_bm_t(-C_{22}^1+C_{23}^1)+km_t(-C_{12}^2+2C_{22}^2-2C_{23}^2) ]
\ \hspace{2cm} \
$$
$$
+a_i[-2v_tm_tC_{22}^3+2a_tm_t(C_{22}^3+C_{23}^3) ] \}
\ \hspace{5.7cm} \
\eqno(a4)
$$
$$
F_{5Z}=\frac{1}{16\pi^2}\sum_{i=\pi, \pi_8, \pi_t}\lambda_i \{
[4v_bm_t(C_{22}^1-C_{23}^1)+km_t(C_{11}^2-C_{12}^2-2C_{22}^2+2C_{23}^2) ]
\hspace{1.5cm} \
$$
$$
+a_i[2a_tm_t(-2C_{11}^3+2C_{12}^3-C_{21}^3-C_{22}^3+2C_{23}^3)+
2v_tm_t(C_{22}^3+C_{23}^3)] \} \ \ \ \
\eqno(a5)
$$
$$
F_{6Z}=\frac{1}{16\pi^2}\sum_{i=\pi, \pi_8, \pi_t}\lambda_i \{
[4v_bm_t(-C_{11}^1+C_{12}^1-C_{21}^1-C_{22}^1+2C_{23}^1)
+km_t(2C_{12}^2+2C_{22}^2)]
$$
$$
+a_i[2v_tm_t(-2C_{11}^3+2C_{12}^3-C_{21}^3-C_{22}^3+2C_{23}^3)+
2a_tm_t(C_{22}^3+C_{23}^3)] \}
\ \hspace{0.2cm} \
\eqno(a6)
$$
where $k=(1 - 2s_W^2)/(2s_Wc_W)$, $v_b=\frac{2}{3}s_W^2/(4s_Wc_W)$,
$a_b=(-1+ \frac{2}{3}s_W^2)/(4s_Wc_W)$;
$B'_i=\frac{\partial}{\partial P_t^2} B_i|_{P_t^2= m_t^2}$;
$C^1=C^1(P_{\bar{t}}$, $-\sqrt{s}$, $m_i$, $m_b$, $m_b)$,
$C^2=C^2$ $(-P_t$, $\sqrt{s}$, $m_b, m_i, m_i)$,
$C^3=C^3$ $(-P_{\bar{t}}$, $-\sqrt{s}$, $m_i, m_t, m_t)$.
The basic two- and three-scalar integral functions can be found in
Ref.\ \cite{Clements83}, and
$$
F_{i \gamma}=F_{i Z}|_{Z \rightarrow \gamma, m_Z=0,
v_t=a_t=1/3, v_b=a_b=-1/6, v_e=a_e=-1/2, k=1}.
\eqno(a7)
$$
For $i=\pi$,
$$
\lambda_{\pi}=(\frac{c_t m_t'}{f_{\pi}})^2, m_t'=\varepsilon m_t,
m_b=0.1m_t', a_i=2.
\eqno(a8)
$$
For $i=\pi_8$,
$$
\lambda _{\pi}=(\frac{m_t' \lambda^a}{f_{\pi}})^2, m_t'=\varepsilon m_t,
m_b=0.1m_t', a_i=2.
\eqno(a9)
$$
And for $i=\pi_t$,
$$
\lambda_t=(\frac{(1-\varepsilon) m_t}{f_{\pi_t}})^2, m_b=4.5 {\rm GeV},
a_i=1.
\eqno(a10)
$$


\begin{thebibliography}{99}

\bibitem{E821}  H.~N~Brown,~{\em et~al.}, Mu g-2 Collaboration,
                  Phys.~Rev.~Lett.{\bf 86}, 2227 (2001).
\bibitem{Weinberg79}
S. Weinberg, Phys. Rev. {\bf D 19}, 1277 (1979);
L. Susskind, Phys. Rev. {\bf D 20}, 2619 (1979).

\bibitem{Dimopoulos79}
S. Dimopoulos and L. Susskind,  Nucl. Phys. {\bf B\ 155}, 237 (1979);
E. Eichten and K. Lane, Phys. Lett. {\bf B 90}, 125 (1980).

\bibitem{Holdom81}
B. Holdom, Phys. Rev.  {\bf D 24}, 1441 (1981);
Phys. Lett. {\bf B 150}, 301 (1985);
T. Appeloquist and D. Karabali, L. C. R. Wijewardhana,
Phys. Rev. Lett. {\bf 57}, 957 (1986);
T. Appeloquist and L. C. R. Wijewardhana, Phys. Rev. {\bf D 36}, 568  (1987);
K. Yamawaki, M. Bando and K. Matumoyo,
Phys. Rev. Lett. {\bf 56}, 1335 (1986);
T. Akiba and T. Yanagida, Phys. Lett. {\bf B 169}, 432 (1986).

\bibitem{Lane89}
K. Lane and E. Eichten, Phys. Lett. {\bf B 222}, 274 (1989);
K. Lane and M. V. Ramana, Phys. Rev.  {\bf D 44}, 2678 (1991).

\bibitem{Chivukula92}
R. S. Chivukula, S. B. Selipsky, and E. H. Simmons,
Phys. Rev. Lett. {\bf 69}, 575 (1992);
R. S. Chivukula, E. H. Simmons, and J.Terning,
Phys. Lett. {\bf B 331}, 383 (1994).  

\bibitem{Miransky89}
V. A. Miransky, M. Tanabashi, and K. Yamawaki,
Phys. Lett. {\bf B 221}, 177 (1989);
W. A. Bardeen, C. T. Hill, and M. Lindner, Phys. Rev. {\bf D 41},
1647 (1990); 
C. T. Hill, D. Kennedy, T. Onogi, H. L. Yu, Phys. Rev. {\bf D 47},
2940 (1993);

\bibitem{Simmons89}
E. H. Simmons, Nucl. Phys. {\bf B 312}, 253 (1989);
C. D. Carone and E. H. Simmons,  Nucl. Phys. {\bf B 397}, 591 (1993);
C. D. Carone and H. Georgi, Phys. Rev. {\bf D 49}, 1427 (1994).

\bibitem{Xiong01}
Z.~Xiong, H.~Chen and L. Lu,  Nucl.~Phys. B{\bf 561},3 (1999);
Z.~Xiong and J.~M.~Yang,  Nucl.~Phys.~B{\bf 602}, 289 (2001),
hep-ph/0012217; Phys. Lett. B{\bf 508}, 295 (2001),  hep-ph/0102259.
\bibitem{Abe95}
F. Abe $et\ al.$, CDF Collaboration, Phys. Rev. Lett. {\bf 74}, 2626(1995);
S. Abachi $et\ al.$, D0 Collaboration, $ibid.$ {\bf 74}, 2632(1995).

\bibitem{Peskin92}
M. E. Peskin, in {\it Physics and Experiments with Linear Collider},
Proceedings of the Workshop, Saarika, Finland, 1991, edited by R. Orava,
P. Eerala, and M. Nordberg (World Scientific, Singapore, 1992), p. 1.

\bibitem{topcref}
C. T. Hill, Phys. Lett. {\bf B 266}, 419 (1991);
S. P. Martin, Phys. Rev. {\bf D 45}, 4283 (1992);
{\it ibid} {\bf D 46}, 2197 (1992); Nucl. Phys. {\bf B 398}, 359 (1993);
M. Lindner and D. Ross, Nucl. Phys. {\bf B 370}, 30 (1992);
R. B\"{o}nisch, Phys. Lett. {\bf B 268}, 394 (1991).

\bibitem{tctwohill}
C. T. Hill, Phys. Lett. {\bf B 345}, 483 (1995).

\bibitem{tctwoklee}
K. Lane and E. Eichten, Phys. Lett. {\bf B 352}, 382 (1995);
K. Lane, Phys. Rev. {\bf D 54}, 2204 (1996);
Phys. Lett. {\bf B 433}, 96 (1998);
G. Lu, Z. Xiong and Y. Cao, Nucl. Phys. {\bf B 487}, 43 (1997); 
Z. Xiao, W. Li, L. Guo and G. Lu, hep-ph/0011175;

\bibitem{Ellis81}
J. Ellis $et\ al.$, Nucl. Phys. {\bf B 182}, 528 (1981);
E. Eichten and K. Lane, Phys. Lett. {\bf B 90}, 125 (1980);
E. Eichten $et\ al.$, Rev. Mod. Phys. {\bf 56}, 579 (1984).

\bibitem{Wu95}
G. H. Wu, Phys. Rev. Lett. {\bf 74}, 4137 (1995);
C. X. Yue $et\ al.$, Phys. Rev. {\bf D 52}, 5314 (1995);
Eur. Phys. J. {\bf C 14}, 313 (2000).

\bibitem{Bohm86}
M. Bohm, W. Hollik and H. Spiesberger, Fortschr Phys. {\bf 34}, 687 (1986);
W. Hollik, $ibid$. {\bf 38}, 165 (1990);
B. Grzad and W. Hollik, Nucl. Phys. {\bf B 384}, 101 (1992).

\bibitem{Yue97}
C. X. Yue $et\ al.$, Phys. Rev. {\bf D 55}, 5541 (1997).

\bibitem{Clements83}
M. Clements $et \ al.$, Phys. Rev. {\bf D 27}, 570 (1983); A. Axelrod, Nucl.
Phys. {\bf B 209}, 349 (1982);
G. Passarino and M. Veltman, Nucl. Phys. {\bf B 160}, 151 (1979).

\end{thebibliography}
\end{document}